\documentclass[12pt]{iopart}
\usepackage{epsfig}
\newcommand{\be}{\begin{equation}}
\newcommand{\ee}{\end{equation}}
\newcommand{\bc}{\begin{center}}
\newcommand{\ec}{\end{center}}
\newcommand{\bea}{\begin{eqnarray}}
\newcommand{\eea}{\end{eqnarray}}
\def\d{{\rm d}}
\begin{document}

\hbox to\hsize{\hfill MPP-2006-28}

\title[Optimal detector locations for SN neutrinos]
{Earth matter effects in supernova neutrinos:
Optimal detector locations}
\author{A.~Mirizzi$^1$, G.G.~Raffelt$^2$
and P.D.~Serpico$^2$}

\address{$^1$~Dipartimento di Fisica and Sezione INFN di Bari,
         Via Amendola 173, 70126 Bari, Italy}
\address{$^2$~Max-Planck-Institut f\"ur Physik
(Werner-Heisenberg-Institut), F\"ohringer Ring 6, 80805 M\"unchen,
Germany}
\eads{alessandro.mirizzi@ba.infn.it,
raffelt@mppmu.mpg.de, and
serpico@mppmu.mpg.de}

\begin{abstract}
A model-independent experimental signature for flavor oscillations in
the neutrino signal from the next Galactic supernova (SN) would be the
observation of Earth matter effects. We calculate the probability for
observing a Galactic SN shadowed by the Earth as a function of the
detector's geographic latitude. This probability depends only mildly
on details of the Galactic SN distribution.  A location at the North
Pole would be optimal with a shadowing probability of about 60\%, but
a far-northern location such as Pyh\"asalmi in Finland, the proposed
site for a large-volume scintillator detector, is almost equivalent
(58\%). We also consider several pairs of detector locations and
calculate the probability that only one of them is shadowed, allowing
a comparison between a shadowed and a direct signal. For the South
Pole combined with Kamioka this probability is almost 75\%, for the
South Pole combined with Pyh\"asalmi it is almost 90\%.  One
particular scenario consists of a large-volume scintillator detector
located in Pyh\"asalmi to measure the geo-neutrino flux in a
continental location and another such detector in Hawaii to measure it
in an oceanic location.  The probability that only one of them is
shadowed exceeds 50\% whereas the probability that at least one is
shadowed is about~80\%. We provide an online tool to calculate
different shadowing probabilities for the one- and two-detector cases.

{\ }

\noindent {\em Keywords}: Supernova neutrinos, neutrino detectors
\end{abstract}
\pacs{97.60.Bw, 
14.60.Pq, 
95.55.Vj 
}

\maketitle

\section{Introduction}\label{introduct}

Galactic core-collapse supernovae are rare, perhaps a few per
century~\cite{Cappellaro:1999qy, Diehl:2006cf}.  However, the large
number of existing or future neutrino detectors with a broad range of
science goals almost guarantees continuous exposure for several
decades, so that a high-statistics supernova (SN) neutrino signal may
eventually be observed. Such a measurement would provide a plethora of
new insights that are crucial both for our astrophysical understanding
of the core-collapse phenomenon~\cite{Woosley:2006ie} and for using
SNe as particle-physics laboratories~\cite{Raffelt:1999tx}. Of
particular interest would be the observation of signatures for flavor
oscillations because this could address a key question of neutrino
physics, i.e.\ whether the neutrino masses are ordered in a normal or
inverted hierarchy~\cite{Dighe:1999bi, Lunardini:2001pb,
Lunardini:2003eh, Dighe:2003be, Takahashi:2002cm, kuo}.

The collapsed core of a SN emits neutrinos and anti-neutrinos of all
flavors with comparable fluxes and spectra~\cite{Jegerlehner:1996kx,
Raffelt:2001kv, Keil:2002in, Raffelt:2003en}. One expects
differences between the spectra and fluxes of $\bar\nu_e$ and
$\bar\nu_{\mu,\tau}$ or between $\nu_e$ and $\nu_{\mu,\tau}$ that
are large enough to observe flavor oscillations, but it will be
difficult to establish such effects solely on the basis of a
$\bar\nu_e$ or $\nu_e$ ``spectral hardening'' relative to
theoretical expectations. Therefore, in the recent literature the
importance of model-independent signatures has been emphasized,
e.g.~in association with the prompt $\nu_e$ neutronization
burst~\cite{Kachelriess:2004ds} or with shock-wave
propagation~\cite{Schirato:2002tg, Tomas:2004gr, Fogli:2004ff,
Fogli:2006xy}. One unequivocal signature would be the observation of
Earth matter effects, because they induce a characteristic
energy-dependent modulation on the measured
flux~\cite{Lunardini:2001pb, Lunardini:2003eh, Dighe:2003be,
Dighe:2003jg, Dighe:2003vm}. In a large-volume liquid scintillator
detector such as the proposed 50~kt Low Energy Neutrino Astronomy
(LENA) project~\cite{Oberauer:2005kw} this modulation could be
detected with high statistical significance~\cite{Dighe:2003jg}.
Moreover, even if single detectors can not resolve these
modulations, comparing the signals from a shadowed with an
unshadowed detector may allow one to diagnose Earth
effects~\cite{Dighe:2003be}.

These intriguing possibilities have motivated us to investigate the
probability for given detector locations to observe the next
Galactic SN in an Earth-shadowed position. Most of the Milky Way is
in the southern sky so that a northern location is obviously
preferred, but a quantitative determination of the probability
distribution of Earth-shadowing as a function of geographic latitude
is missing, except for a brief discussion in
Ref.~\cite{Lunardini:2001pb}. An additional motivation for our work
is that large-volume scintillator detectors are being discussed for
the purpose of geo-neutrino observations.  After KamLAND's
pioneering measurement of the $\bar\nu_e$ flux from uranium and
thorium $\beta$-decays in the Earth~\cite{Araki:2005qa}, performed
in a complex geophysical environment with a large $\bar\nu_e$
background from power reactors, the exploration of different
detector sites has become crucial.  The choice of location could be
influenced by the role of such detectors as excellent SN neutrino
observatories, in particular because they need to operate for a long
time to accumulate meaningful statistics for geo-neutrino studies.
For example, if a large-volume scintillator detector were built in
the Pyh\"asalmi mine (Finland)~\cite{Finland} to measure the
geo-neutrino flux mainly from the continental crust and another one
in Hawaii~\cite{hanohano} to measure neutrinos from the oceanic
crust, the ``geographic complementarity'' of these locations with
regard to SN shadowing is important.

We begin in Sec.~\ref{sec:SNdistribution} with a discussion of the
distribution of core-collapse SNe in the Galaxy and then determine
the SN probability distribution in the sky. In
Sec.~\ref{sec:detectorlocation} we determine the probability for
Earth shadowing as a function of geographic location of one or two
detectors. We also provide an online tool for the
 calculation of these probabilities~\cite{webshadowing}.
We conclude our work in Sec.~\ref{sec:conclusions}.

\section{Supernova distribution in the Milky Way}
                                            \label{sec:SNdistribution}
In this section we characterize the SN probability distribution in
our Galaxy. In Sec.~\ref{sec:model} we present the models adopted
for the SN volume distribution. In Sec.~\ref{sec:proj} we discuss
the probability of a core-collapse SN in terms of Galactic and
Equatorial coordinates. Finally in Sec.~\ref{sec:dist} we estimate
the distance distribution of SN events. The SN distribution in the
Galaxy is expressed in cylindrical galactocentric coordinates
$(r,z,\theta)$, where the origin corresponds to the Galactic center,
$r$ indicates the radial coordinate, $\theta$ the azimuthal angle
and $z$ the height with respect to the Galactic plane.

\subsection{Models for the volume distribution}
\label{sec:model}

The probability distribution of core-collapse SNe in the Milky Way is
not well known. These SNe mark the final evolution of massive stars
and thus must occur in regions of active star formation, i.e.\ in the
Galactic spiral arms. As proxies for the core-collapse SN distribution
one can use either observations of other Galaxies, or in our Galaxy
the distribution of pulsars and SN remnants (SNRs), the distribution
of molecular hydrogen (H$_2$) and ionized hydrogen (HII) and the
distribution of OB-star formation regions. (For a review see
Ref.~\cite{Ferriere:2001rg} and references there.) These observables
are either directly connected with core-collapse events (SNRs,
pulsars) or with young, massive star formation activity and the
related emission of ultraviolet light (H$_2$, HII, OB stars). All of
these observables are consistent with a deficit of SNe in the inner
Galaxy and a maximum of the probability at 3.0--5.5~kpc galactocentric
distance.

The star formation activity is not smoothly distributed in the
Galaxy. Besides generally following the spiral arms, it can be
concentrated in small regions or spike-like complexes like the Cygnus
OB association.  Small regions of high star-forming activity have also
been found within 50~pc from the Galactic center~\cite{Figer:2003tu}
that may contribute up to 1\% of the total Galactic star formation
rate, although this finding does not seem to contradict the overall
picture of a reduced SN rate in the inner Galaxy.

However, we are only interested in the SN distribution
projected on the sky and the Earth's rotation introduces an
additional averaging effect. Therefore,
it will be enough to consider a smooth
distribution with azimuthal symmetry. In particular, we shall use the
following common parametrization for the Galactic surface density of
core-collapse (cc) events,
\begin{equation}\label{param}
{\sigma}_{\rm cc}(r)\propto r^\xi\exp(-r/u)\ ,
\end{equation}
where $r$ is the galactocentric radius.  For the birth location of
neutron stars, a fiducial distribution of this form was suggested with
the parameters~\cite{Yusifov:2004fr}
\begin{equation}\label{eq:par2}
\textrm{Neutron~stars:}\qquad\left\{
\begin{array}{l}
  \xi = 4 \,,\nonumber\\
    u = 1.25~{\rm kpc}\,. \\
\end{array}\right.
\end{equation}
These parameters are consistent with several SN-related observables,
even though large uncertainties remain~\cite{Yusifov:2004fr}.  On the
other hand, the pulsar distribution indicates~\cite{Lorimer:2003qc}
\begin{equation}\label{eq:par1}
\textrm{Pulsars:}\qquad\left\{
\begin{array}{l}
  \xi = 2.35\,,\nonumber\\
u = 1.528~{\rm kpc}\,.
\end{array}\right.
\end{equation}
We will use the parameters of Eq.~(\ref{eq:par2}) as our benchmark
values and those of Eq.~(\ref{eq:par1}) as an alternative model to
illustrate the dependence of our results on different input
choices. 

\begin{figure}[!t]\bc
\epsfig{file=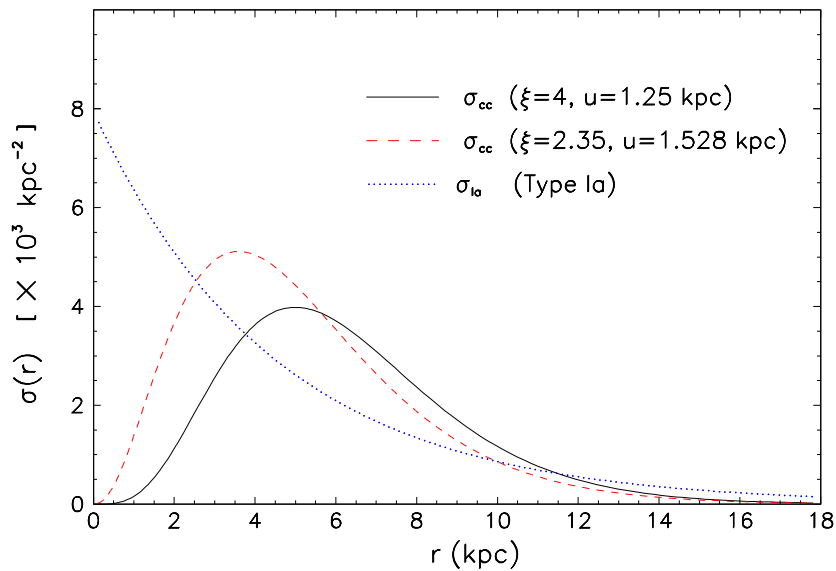,width=0.7\columnwidth}\ec \caption{Galactic
surface density of core-collapse SNe as a function of galactocentric
radius according to Eq.~(\ref{param}).  {\em Solid curve:} Our
benchmark case with the parameters of Eq.~(\ref{eq:par2}). {\em
Dashed curve:} Alternative parameters of Eq.~(\ref{eq:par1}). {\em
Dotted curve:} Type~Ia SNe according to Eq.~(\ref{distrTIa}). The
surface densities are normalized as $\int{\sigma}(r)\,2\pi\,r\d
r=1$. \label{radial}}
\end{figure}

In Fig.~\ref{radial} we show the normalized Galactic surface density
of core-collapse SNe according to Eq.~(\ref{param}) as a function of
the galactocentric distance for the two choices of parameters given in
Eqs.~(\ref{eq:par2}) and~(\ref{eq:par1}). Our benchmark distribution
shows a peak at $r=5$~kpc, while for the parameters of
Eq.~(\ref{eq:par1}) the surface density peaks at a lower distance
$r\simeq 3$~kpc.  The distributions are normalized as $\int \d r \,
{\sigma}(r)\,2\pi\,r=1$, i.e.\ the surface densities $\sigma$ are
given in SNe per kpc$^2$. Of course, to obtain the number of SN events
per Galactic unit surface and unit time, the quantities $\sigma$ would
have to be multiplied with the integrated Galactic SN rate. However,
since we are only interested in the relative probability of SNe in
different regions of the sky, the overall rate is not important for
our study.

The vertical distribution of neutron stars at birth with respect to
the Galactic plane can be approximated by the superposition of a thin
Gaussian disk with a scale height of 212~pc and a thick disk with
three times this scale height, containing 55\% and 45\% of the
pulsars, respectively~\cite{Ferriere:2001rg},
\begin{equation}
{\cal R}_{\rm cc}(z)\propto 0.79 \ \exp \left[ - \left( {z \over 212
\ {\rm pc}} \right)^2 \right] + 0.21 \ \exp \left[ - \left( {z \over
636 \ {\rm pc}} \right)^2 \right],
\end{equation}
where $z$ is the height above the Galactic plane. We assume this
vertical distribution to be independent of galactocentric
distance~\cite{Lorimer:2003qc} so that
\begin{equation}
n_{\rm cc}(r,z)\propto \sigma_{\rm cc}(r)\, {\cal R}_{\rm
cc}(z)
\label{sn2volprob}
\end{equation}
is the volume distribution.

We stress that the distribution of core-collapse SNe differs
significantly from the overall matter distribution, particularly in
the inner part of the Galaxy. The distribution of Type Ia SNe---that
are believed to originate from old stars in binary systems---more
closely follows the matter distribution. It can be parameterized
as~\cite{Ferriere:2001rg}
\begin{equation}\label{distrTIa}
n_{\rm Ia}(r,z)={\sigma}_{\rm Ia}(r){\cal R}_{\rm Ia}(z) \propto
\exp\left( - \frac{r}{4.5~{\rm kpc}} \right)
\exp \left(-\frac{|z|}{325~{\rm pc}} \right)\ .
\end{equation}
In Fig.~\ref{radial} we show the monotonically falling SNe~Ia
surface density $\sigma_{\rm Ia}(r)$ for comparison with the
core-collapse case.

\subsection{Projection on the sky}                   \label{sec:proj}

The probability of a core-collapse SN as a function of the Galactic
longitude $l$ and latitude $b$ is given by an integration along the
line of sight,
\begin{equation}\label{Plb}
P(l,b)\propto \int_0^\infty \d s\, n_{\rm cc}[r(s,l,b),z(s,b)]\ ,
\end{equation}
where
\begin{eqnarray}
r&=&\left(s^2\cos^2b+d_\odot^2-2\,s\,d_{\odot}\cos l\cos b
\right)^{1/2}\ ,\nonumber\\
z&=&s\,\sin b\ .
\end{eqnarray}
Here, $- \pi \leq l \leq \pi$, $-\pi/2 \leq b \leq \pi/2$, and
$d_\odot\simeq 8.5$~kpc is the solar distance from the Galactic
center. The function $P(l,b)$ can be recast in terms of the
Equatorial coordinates ($\alpha,\delta$), in which the reference
plane is the Earth equator instead of the Galactic plane. The
transformation relating the two systems of coordinates is
\begin{eqnarray}
\sin \delta& = \sin b \sin \delta_{\rm NGP} +
\cos b \cos \delta_{\rm NGP} \sin (l - l_0) \,\ , \\
\cos(\alpha - \alpha_0) &= \cos (l-l_0) \cos b / \cos \delta \,\ , \\
\sin(\alpha - \alpha_0) &= [- \sin b \cos \delta_{\rm NGP} + \cos b
\sin \delta_{\rm NGP}
 \sin (l-l_0)]/\cos \delta \,\ ,
\end{eqnarray}
where, for the Julian epoch J2000, the coordinates of the north
Galactic pole (NGP) are $\alpha_{\rm NGP} = 12~{\rm h}~51.42~{\rm
m}$ and $\delta_{\rm NGP} = 27^{\circ} 07.8^{\prime}$. Hence, the
ascending node of the Galactic equator is at $\alpha_0
=282.86^{\circ}$ and $l_0 = 32.93^{\circ}$. Further details on
astronomical coordinate systems can be found, for example,
in Ref.~\cite{Lang}.

\begin{figure}[!htb]
\bc
\begin{tabular}{c}
\epsfig{file=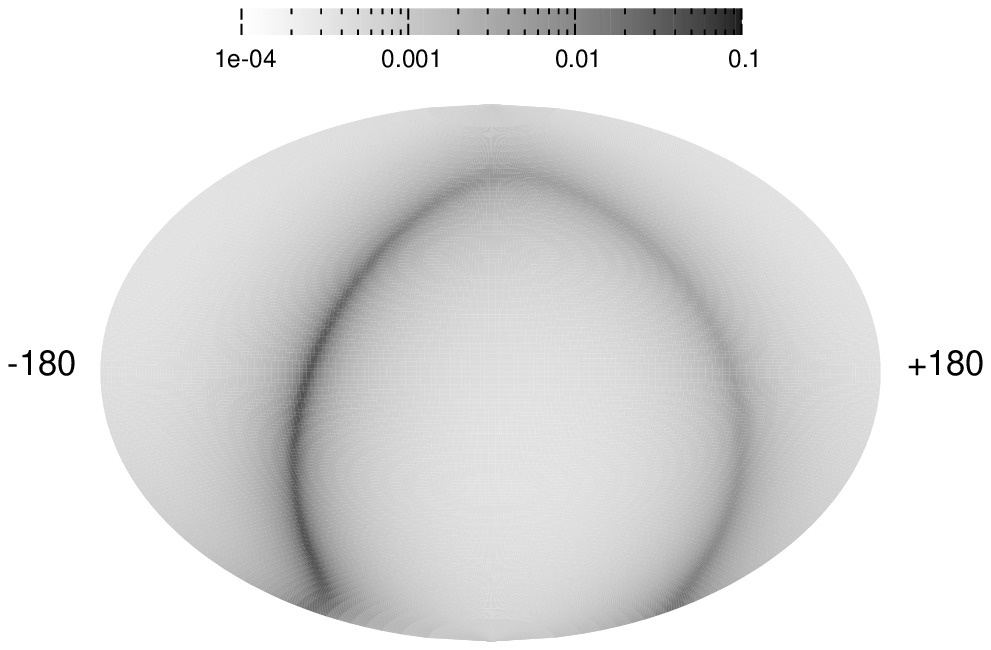,width=0.7\columnwidth}\\[-1cm]
\epsfig{file=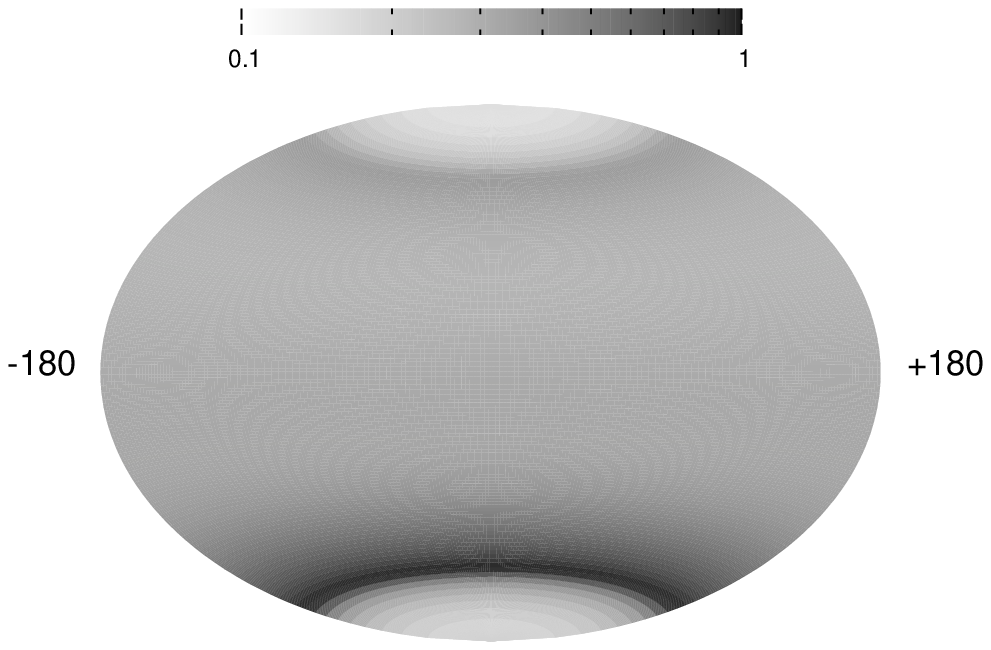,width=0.7\columnwidth}
\end{tabular}
\vskip-1cm
\caption{Probability distribution $P(\alpha, \delta)$
of core-collapse SNe in the sky. {\em Top:}~Earth Equatorial
coordinates. {\em Bottom:}~Average over right-ascension, see
Eq.~(\ref{eq:omega}).} \label{figmap} \ec
\end{figure}
The sky map in Equatorial coordinates is shown in the upper panel of
Fig.~\ref{figmap} for our benchmark distribution. However, since the
neutrino detectors are fixed to the Earth, we only need this
distribution averaged over time, i.e.~over right ascension
$\alpha$. Therefore, all relevant information is contained in the
``exposure probability function''
\begin{equation}
\omega (\delta)\propto\int_{-\pi}^{+\pi}
\d\alpha\,P(\alpha,\delta)\,.
\label{eq:omega}
\end{equation}
It provides the probability distribution of the arrival direction of
a SN signal in terms of declination. We show the time-averaged sky
map in the bottom panel of Fig.~\ref{figmap}, normalized as $\int\d
\delta\, \omega(\delta)\cos\delta=1$. This normalization also fixes
the overall constant for the function $P(\alpha, \delta)$ plotted in
the top panel of Fig.~\ref{figmap}. Fig.~\ref{figmap} clearly shows
the preference of southern locations in the sky and also shows the
polar regions that are nearly void of SNe because the Milky Way
extends approximately between declinations of $-60^\circ$ to
$+60^\circ$.

\begin{figure}
\bc
\epsfig{file=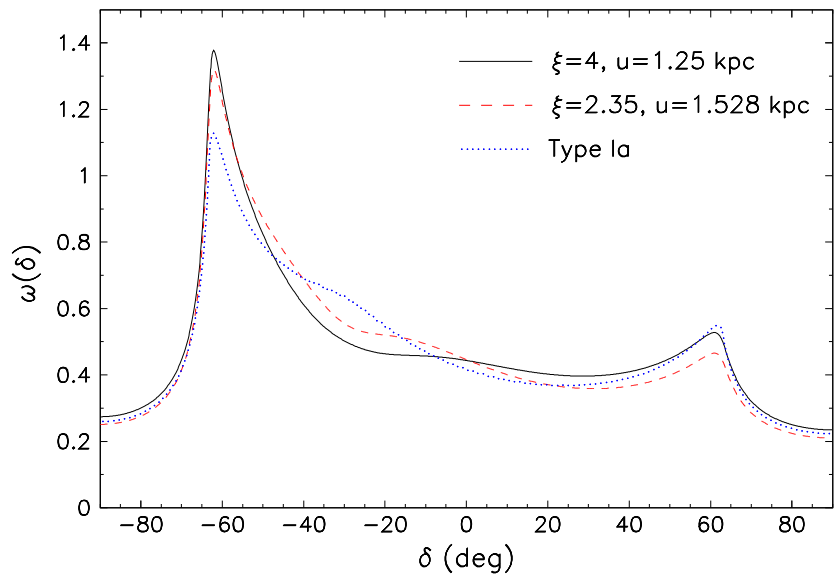,width=0.7\columnwidth}
\ec
\caption{The ``exposure function'' $\omega(\delta)$ for different
  Galactic SNe distributions. {\em Solid curve:} Benchmark
  model based on Eq.~(\ref{param}) with the parameters of
  Eq.~(\ref{eq:par2}). {\em Dashed curve:} Same with the
  alternative parameters of Eq.~(\ref{eq:par1}).  {\em Dotted curve:}
 Distribution of SNe~Ia for
  comparison.\label{robust}}
\end{figure}

In Fig.~\ref{robust} we show the exposure probability function
$\omega(\delta)$ for different assumptions about the Galactic SN
distribution, i.e.~our benchmark distribution Eq.~(\ref{param}) with
the parameters of Eq.~(\ref{eq:par2}), the alternative parameters of
Eq.~(\ref{eq:par1}), and as an extreme case the SN~Ia distribution of
Eq.~(\ref{distrTIa}) which, of course, is not realistic for
core-collapse SNe. The normalization $\int\d \delta\,
\omega(\delta)\cos\delta=1$ is adopted throughout. In this
normalization of $\omega(\delta)$ an isotropic SN distribution would
correspond to a horizontal line in Fig.~\ref{robust}.  The exposure
function depends only mildly on details of the assumed Galactic SN
distribution, because the geometric effect dominates that the SN
probability is negligible outside the Galactic disk. The largest model
variation occurs around a declination of $-30^\circ$ near the Galactic
center region.  The edges at about $\pm60^\circ$ correspond to the
region where the Milky Way would end in the sky if it were infinitely
thin. The tails beyond these declinations come from the vertical
extension of the Galactic disk around us where we have assumed the
solar system to be located exactly in the Galactic plane.

\subsection{Distance distribution}
\label{sec:dist}

\begin{figure}
\bc
\epsfig{file=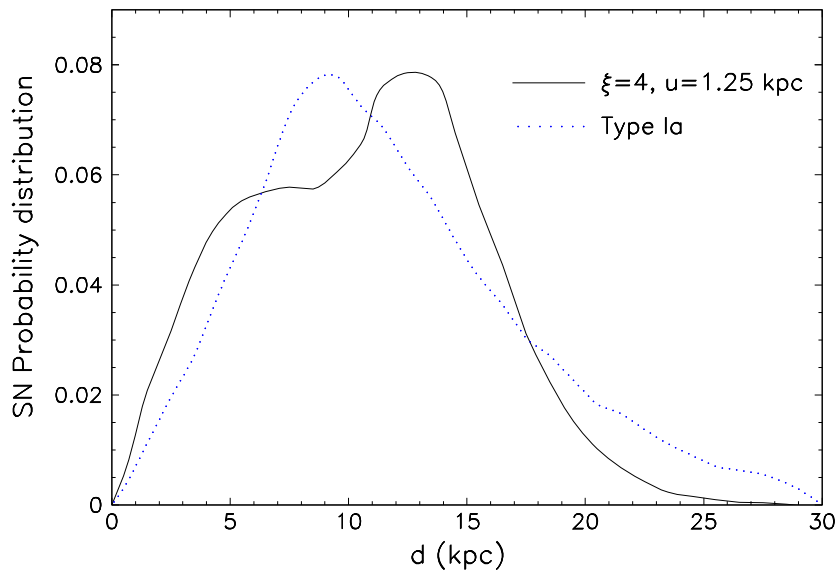,width=0.7\columnwidth}
\ec
\caption{SN probability vs.\ distance from the Sun. {\em Solid curve:}
Benchmark model based on Eqs.~(\ref{param}) and~(\ref{eq:par2}).  {\em
Dotted curve:} SNe~Ia assuming
Eq.~(\ref{distrTIa}).\label{distribution}}
\end{figure}

As an aside, we use our assumed core-collapse SN distribution in the
Galaxy to evaluate the SN distance distribution relative to the solar
system.  The average distance is
\begin{equation}
\langle d_{\rm cc}\rangle=\frac{\int n_{\rm
cc}(r,z)d(r,z,\theta)\,r\, \d r \d z \d \theta }{\int n_{\rm
cc}(r,z)2\pi r\d r\d z}\label{avdII},
\end{equation}
where
\begin{eqnarray}
d(r,z,\theta)&=&
\left[(x-x_\odot)^2+(y-y_\odot)^2+(z-z_\odot)^2\right]^{1/2}
\nonumber\\
&=&\left[r^2+z^2+d_\odot^2-2r d_\odot\cos\theta\right]^{1/2}\,.
\end{eqnarray}
For the distribution $n_{\rm cc}$ of Eq.~(\ref{sn2volprob}) with our
benchmark parameters of Eq.~(\ref{eq:par2}) we find $\langle d_{\rm
cc}\rangle=10.7~{\rm kpc}$ with a rms dispersion of 4.9~kpc. While
the average distance agrees with the fiducial distance of 10~kpc
that is frequently assumed in the literature, we note that the
dispersion of distances is very large. This is clearly visible in
Fig.~\ref{distribution} where we plot the probability distribution
$\Pi(d)$ of SN events as a function of the distance $d$ from the
Sun, so that $\int_a^b \Pi(x) \d x$ gives the probability that a SN
happens between distance $a$ and $b$ from the Sun.
 The dip in
the middle corresponds to the deficit of core-collapse SNe in the
Galactic center region.  For comparison we also show the probability
distribution for SNe~Ia according to Eq.~(\ref{distrTIa}). Their
average distance is 11.9~kpc with a rms dispersion of 6.0~kpc.
On the other hand,  the median distances are quite similar for
the two distributions, and equal to about 10.9~kpc.
 The
large-distance behavior for both cases differs significantly due to
the different scales in the exponential tails of Eqs.~(\ref{param})
and~(\ref{distrTIa}). Of course, the scarcity of data at large
galactocentric distances implies that the extrapolation may be
unphysical.

\section{Optimal detector location}       \label{sec:detectorlocation}

\subsection{One detector}\label{onedet}

Armed with the exposure function $\omega(\delta)$ shown in
Fig.~\ref{robust} we now determine the probability that a detector
located at a geographic latitude $\lambda$ will observe the next
Galactic SN below the Earth horizon. Of course, observable matter
effects would require a minimal path-length of a few thousand
kilometers~\cite{Dighe:2003jg} so that our Earth-shadowing criterion
is somewhat schematic. Moreover, the phenomenological signature of
Earth matter effects also depends on whether or not the neutrinos
cross the core~\cite{Dighe:2003vm, Petcov:1998su, Akhmedov:1998ui}.
We use a core radius $R_{\rm c}=3486$~km and an Earth radius of
$R_\oplus=6371$~km~\cite{Earthdata}. The Earth or core shadowing
condition for a source with altitude $a$ with respect to the
horizon~is
\begin{equation}\label{eq:crossing}
\sin a<\kappa=\cases{0 \,\ , &Earth shadowing\ ,\cr
 -\sin a_{\rm c} \,\ ,&Core shadowing\ ,}
\end{equation}
where
\begin{equation}
\sin a_{\rm c}=\sqrt{1-(R_{\rm c}/R_\oplus)^2}=0.837\ .
\end{equation}
In general, for a  neutrino path-length $L$ in the Earth, one has
\begin{equation}
\kappa = - \sin a_L =-\frac{L}{2 R_\oplus}.
\end{equation}
The altitude of an object of equatorial coordinates $(\alpha,\delta)$ is
\begin{equation}\label{altitude}
\sin a=\sin\lambda\sin\delta+\cos\lambda\cos\delta\cos H\ ,
\end{equation}
where $H=t-\alpha$ is the hour angle and $t$ the local sidereal
time. Since no time information on the next SN is available in
advance, $H$ has to be taken  as a random variable.

For polar locations ($\lambda = \pm~\pi/2$) the shadowing condition
simplifies, because $\cos\lambda=0$ and the time dependence
disappears. The geometrical probability $p_\kappa(\lambda,\delta)$
for a trajectory to satisfy one of the  shadowing conditions is
\begin{equation}
p_\kappa\left(\pm~\frac{\pi}{2},\delta\right)
=\Theta(\kappa\mp\sin\delta)\label{pmcond}\ ,
\end{equation}
where $\Theta(x)$ is the usual step function.  A similar
simplification holds for objects located at the celestial poles
($\delta=\pm~\pi/2$) for which $\sin\,a=\pm \sin\lambda$ and
\begin{equation}\label{ppoles}
p_\kappa\left(\lambda,\pm~\frac{\pi}{2}\right)=
\Theta(\kappa\mp\sin\lambda)\ .
\end{equation}
Equation~(\ref{ppoles}) is related to Eq.~(\ref{pmcond}) by the
symmetry $p_\kappa({\lambda},\delta)=p_\kappa(\delta,\lambda)$.  This
is a general property following directly from Eq.~(\ref{altitude}).

Apart from these special cases, we have both $\cos\lambda>0$ and
$\cos\delta>0$ so that the general shadowing condition
Eq.~(\ref{eq:crossing}) can be recast as
\begin{equation}
\cos H<-\tan\lambda\tan\delta
+\frac{\kappa}{\cos\lambda\cos\delta}\ ,\label{condition0}
\end{equation}
or equivalently
\begin{equation}
\cos H>\tan\lambda\tan\delta
-\frac{\kappa}{\cos\lambda\cos\delta}\equiv
h_\kappa(\lambda,\delta)\ .\label{condition}
\end{equation}
Note that we have replaced $H+ \pi$ with $H$ in the argument of the
cosine, since both are random variables. Unless
Eq.~(\ref{condition}) is always or never satisfied, one has
\begin{equation}
-\arccos h_\kappa(\lambda,\delta)<H({\rm mod}~2\pi)<\arccos
h_\kappa(\lambda,\delta) \,\ .
\end{equation}
The general solution is then
\begin{equation}\label{cond2}
p_\kappa(\lambda,\delta)=
\cases{
\Theta[1-h_\kappa(\lambda,\delta)]\,\Theta[-h_\kappa(\lambda,\delta)-1]
&for $|h_\kappa(\lambda,\delta)|\geq 1$,\cr
\frac{1}{\pi}\arccos\,h_\kappa(\lambda,\delta)&otherwise.}
\end{equation}
This follows because all values of the random variable $H({\rm
mod}\:2\pi)$ in the interval $(-\pi,\pi]$ are equally likely. As a
check of Eq.~(\ref{cond2}) we note that for a detector at the
equator ($\lambda=0$) it simplifies to
\begin{equation}
p_\kappa(0,\delta)=
\frac{1}{\pi}\arccos\left(\frac{\kappa}{\cos\delta}\right)\ ,
\end{equation}
reproducing the trivial result $p_0(0,\delta)=1/2$.

In order to obtain the Earth shadowing probability for the neutrinos from
the next Galactic SN, given the detector latitude $\lambda$, we must
convolve with the distribution of source declination angles, i.e.
with the exposure function, so that
\begin{equation}
P_\kappa(\lambda)=\int \d \delta \cos\delta\,\,
p_\kappa(\lambda,\delta)\,\omega(\delta)\ ,
\end{equation}
where we have used the normalization $\int\d \delta
\cos\delta\,\omega(\delta)=1$.

\begin{figure}
\bc
\epsfig{file=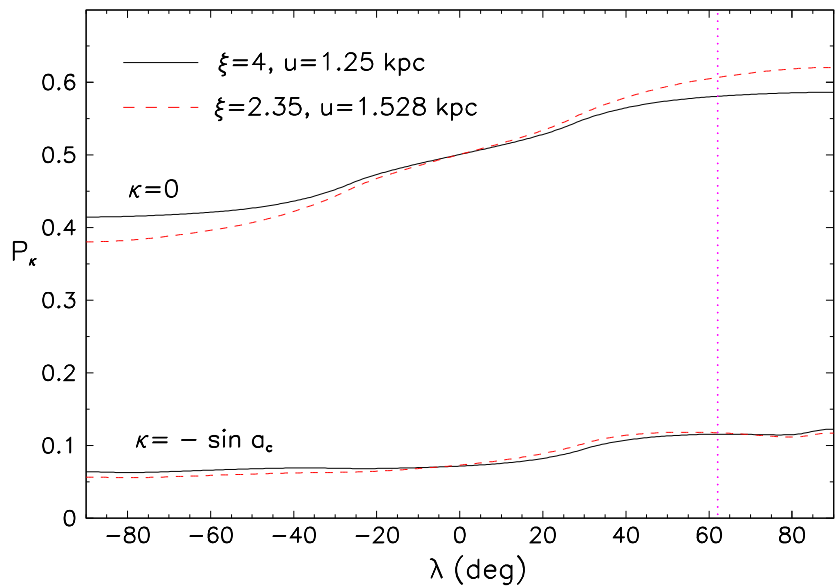,width=0.7\columnwidth}
\ec
\caption{Probability for Earth shadowing (upper curves) and core
shadowing (lower curves).  {\em Solid lines:} Galactic
benchmark distribution with parameters given by Eq.~(\ref{eq:par2}).  {\em
Dashed lines:} Alternative parameters of Eq.~(\ref{eq:par1}).  The
latitude of the Pyh{\"a}salmi site is indicated by the vertical dotted
line.}\label{next}
\end{figure}

We show the Earth and core shadowing probability as a function of
detector latitude in Fig.~\ref{next}. The solid lines refer to our
Galactic benchmark distribution Eq.~(\ref{param}) with parameters
Eq.~(\ref{eq:par2}) whereas the dashed lines refer to the
alternative parameters Eq.~(\ref{eq:par1}). Once more we find that
the dependence on details of the Galactic distribution is mild. Note
that for the Earth-shadowing case $P_0(\lambda)+P_0(-\lambda)=1$
because all information on the longitude is lost and two locations
at $\lambda$ and $-\lambda$ are complementary: if one detector is
shadowed, the antipodal one is certain not to be shadowed.

The largest probability for Earth  shadowing is a the North pole,
but the Pyh{\"a}salmi site in Finland is almost equivalent. The
behavior for the core shadowing condition is similar, although the
advantage of a northern site is somewhat boosted: for $\kappa= -\sin
a_c$, the probability $P_{\kappa}(\lambda)$ varies by almost a
factor two moving from a far-southern location to a far-northern
one, where it reaches almost 12\%. If one asks for a minimal
neutrino path-length of $L=3000$~km~\cite{Dighe:2003jg}, the
shadowing probability is about 10\% less than the shown
Earth-crossing case. However, the minimum path-length to detect
Earth signatures depends on the detailed features of the
flavor-dependent neutrino fluxes and on the detector properties.
Therefore, to keep our discussion simple and general we restrict
ourselves to the illustrative cases of Earth and core crossing.
Note that the generic case can be calculated with the online
tool~\cite{webshadowing}.

Some representative locations are tabulated in
Table~\ref{tab:loctab}. The list is not exhaustive, but rather meant
to provide representative and complementary geographic locations. It
includes two existing locations of high-statistics experiments
sensitive to SN neutrinos, i.e.~Super-Kamiokande in
Japan~\cite{SKde} and IceCube that is under construction at the
South Pole~\cite{Ahrens:2002dv}, and a few possible locations of
next-generation detectors proposed in the literature. In particular,
the possibility of a Megatonne water-Cherenkov detector is discussed
worldwide, including the Underground nucleon decay and Neutrino
Observatory (UNO) with a possible location in the Soudan mine
(Minnesota, USA)~\cite{Jung}, the Hyper-Kamiokande (HK) detector in
the Tochibora Mine (Kamioka region, Japan) \cite{HK03}, and the
MEgaton class PHYSics (MEMPHYS) detector at the Fr{\'e}jus site
(France-Italy)~\cite{Campagne:2006yx}. Different locations are under
study for the large-volume scintillator detector LENA. The
Pyh\"asalmi mine in Finland and Hawaii are of interest for
geo-neutrino studies. Hawaii and Australia have also been discussed
as favored sites for detecting the cosmic diffuse SN neutrino
background because of the low reactor background~\cite{nove}.

\begin{table}
\caption{\label{tab:loctab}Representative locations of proposed or existing
SN neutrino detectors and neutrino shadowing probabilities,
assuming our benchmark Galactic distribution.}
\medskip
\hbox{\hskip6em\begin{tabular}{llclccc}
\hline
Location &\multicolumn{2}{l}{Latitude}&
\multicolumn{2}{l}{Longitude}&
\multicolumn{2}{l}{Shadowing probability}\\
&&&&&Earth&Core\\
\hline
Pyh{\"a}salmi, Finland & 63.66$^\circ$&N&  26.04$^\circ$&E& 0.581 & 0.116\\
Soudan, USA            & 47.82$^\circ$&N&  92.23$^\circ$&W& 0.572 & 0.112\\
Fr\'ejus, France-Italy & 43.43$^\circ$&N&   6.73$^\circ$&E& 0.568 & 0.110\\
Kamioka, Japan         & 36.27$^\circ$&N& 137.3$^\circ$ &E& 0.560 & 0.104\\
Hawaii, USA            & 19.70$^\circ$&N& 156.30$^\circ$&W& 0.528 & 0.082\\
Sydney, Australia      & 33.87$^\circ$&S& 151.22$^\circ$&E& 0.445 & 0.069\\
South Pole             & 90$^\circ$   &S& ---&&             0.414 & 0.064\\
\hline
\end{tabular}}
\end{table}

\subsection{Two detectors}\label{twodet}

Next we consider the simultaneous SN neutrino detection by two
different detectors. A single detector can observe Earth matter
effects only if it has good energy resolution and, of course, enough
statistics. A scintillator detector like the 50-kton LENA project
has good energy resolution and can detect unambiguously these
signatures, whereas water Cherenkov detectors have an intrinsically
poorer energy resolution so that Megaton-scale masses would be
necessary~\cite{Dighe:2003jg}. However, smaller detectors, even if
they can not resolve directly the modulations, could still reveal
Earth effects by the comparison of the signals from a shadowed with
an unshadowed detector. Of particular interest are Super-Kamiokande
in Japan and IceCube at the South Pole that will be completed in a
few years. In this case the Earth effect will be detected as a
difference in the signal normalization between the two
detectors~\cite{Dighe:2003be}. To this end it would be crucial that
one of the detectors is shadowed whereas the other is not, i.e.\ we
are interested in the probability for exactly one of them to be
shadowed.

The special location of IceCube makes this kind of problem a very
simple generalization of the previous one-detector case. The
shadowing condition of Eq.~(\ref{eq:crossing}) is satisfied or
missed at the South Pole with a probability given by the lower or
upper sign of Eq.~(\ref{pmcond}), and at any other location
according to Eq.~(\ref{cond2}) or its complement to 1. Note that we
can safely use the formulae of Sec.~\ref{onedet} since the location
of one detector at the South Pole implies that no longitude
difference enters the problem.

In the general case of two detectors at geographic latitudes
$\lambda$ and $\lambda'$ with a difference in longitude of $\Delta$,
the altitudes $a$ and $a'$ of an object located at ($\alpha,\delta$)
are
\begin{equation}
\left\{
\begin{array}{ll}
\sin\,a=\sin\lambda\sin\delta+\cos\lambda\cos\delta\cos H \,\ ,\\
\sin\,a'=\sin\lambda'\sin\delta+\cos\lambda'\cos\delta\cos(H+\Delta)\,\ .
\end{array} \right.
\end{equation}
To maximize the chance to detect Earth matter effects, the most
interesting case for two general detectors is that exactly one of
them sees a SN from a shadowed position, thus allowing one to
compare a shadowed signal with an unshadowed one. The probability
for detector I to be shadowed and detector II not is: (i)~zero if
one of the two probabilities vanishes when calculated as described
in the previous section; (ii)~trivially equal to the one-detector
case when the probability of one of the single-detector conditions
is 1; (iii)~in all remaining cases it can be evaluated from the
following system of inequalities where we omit $({\rm mod}\:2\pi)$
\begin{equation}
\left\{
\begin{array}{ll}
-\arccos h_\kappa(\lambda,\delta)<H<\arccos h_\kappa(\lambda,\delta)\ ,\\
\arccos h_\kappa({\lambda'},\delta)<H+\Delta<2\pi-\arccos
h_\kappa({\lambda'},\delta)\ .
\end{array} \right.
\end{equation}
The fraction of the $2\pi$-angle satisfying the previous conditions
is the desired probability. The complementary situation,
i.e.~detector II is shadowed and detector I not, is
\begin{equation}
\left\{
\begin{array}{ll}
\arccos h_\kappa(\lambda,\delta) <H
<2\pi-\arccos h_\kappa(\lambda,\delta)\ ,\\
-\arccos h_\kappa({\lambda'},\delta)<H+\Delta<\arccos
h_\kappa({\lambda'},\delta)\ ,
\end{array} \right.
\end{equation}
apart for the trivial cases (i) and (ii) described above.

It is also worthwhile to consider the situation that two detectors
with excellent energy resolution are available, e.g.~two
scintillator detectors. We have already mentioned that
next-generation scintillator detectors may be built at two different
sites, e.g.\ one in Europe and another in Hawaii, for the purpose of
geo-neutrino research. In this case, the most interesting case is
that at least one detector sees a shadowed signal. The
generalization of the previous formulae to this case is
straightforward.

Our results are reported in Table~\ref{tab:twodetect} for the
Earth-crossing case and in Table~\ref{tab:twodetectK} for core
crossing. The quantities listed in each cell at row $i$ and column
$j$ are the following probabilities:
\begin{eqnarray}\label{eq:cellentries}
P(i,\bar{j})&\quad&\hbox{Detector $i$ shadowed, $j$ not shadowed}\ ;\nonumber\\
P(\bar{i},{j})&\quad&\hbox{Detector $i$ not shadowed, $j$
shadowed}\ ; \nonumber\\
P(i,j)&\quad&\hbox{Both detectors shadowed}\ .
\end{eqnarray}
For example, in Table~\ref{tab:twodetect} in the row ``Pyh\"asalmi''
and  column ``South Pole'' we read that the probability that a SN
will be shadowed at Pyh\"asalmi but not at the South Pole is 0.519,
for the reverse case is 0.353, whereas the chance that both
detectors see it shadowed is 0.062.

\begin{table}[b]
\caption{Shadowing probability for two detectors. The quantities
reported in each cell are explained in Eq.~(\ref{eq:cellentries}).}
\label{tab:twodetect}
\begin{center}
\begin{tabular}{|c||c|c|c|c|c|c|} \hline
LOCATIONS     & South Pole & Sydney & Hawaii & Kamioka & Fr\'ejus  & Soudan \\
\hline
\hline
              &    .519    &  .457  &  .285   & .179   &  .065    &   .148 \\
Pyh{\"a}salmi &    .353    &  .315  &  .231   & .157  &   .052    &   .139 \\
              &    .062    &  .130  &  .296   & .400  &    .516   &   .433 \\
\cline{1-7}
              &  .475      &  .437   &  .187  &   .221  &  .162            \\
Soudan        &  .317      &  .310   &  .142  &   .208   & .158            \\
              &  .097      &  .135   &  .386  &   .351   &  .410           \\
\cline{1-6}
              &  .461      &  .484   &   .326     &  .230                  \\
Fr\'ejus      &  .307      &  .361   &   .285     &  .220                  \\
              &  .107      &  .084   &   .242     &  .339                  \\
\cline{1-5}
              &   .435     &   .278     &  .184                      \\
Kamioka       &   .290     &   .164     &  .152                     \\
              &    .124    &   .281    &   .375                    \\
\cline{1-4}
              &      .362      &   .251                              \\
Hawaii        &      .249      &   .168                           \\
              &      .165      &   .277                          \\
\cline{1-3}
              &  .160                            \\
Sydney        &  .129                            \\
              &  .285                            \\
\cline{1-2}
\end{tabular}
\end{center}
\end{table}

\begin{table}[htb]
\caption{Same as Table~\ref{tab:twodetect} for core
shadowing.}\label{tab:twodetectK}
\begin{center}
\begin{tabular}{|c||c|c|c|c|c|c|} \hline
LOCATIONS     & South Pole & Sydney & Hawaii & Kamioka & Fr\'ejus  & Soudan \\
 \hline \hline
              &   .115    &   .116  &  .115  &  .116    &  .041    &  .110  \\
Pyh{\"a}salmi &   .064    &   .069  &  .082  &  .114    &  .035    &  .107  \\
              &   .000    &   .000  &  .000  &  .000    &  .074    &  .005  \\
\cline{1-7}
              &   .112   &   .112     &   .107     &    .112    &  .111     \\
Soudan        &   .064   &   .069     &   .077     &    .104    &  .108     \\
              &   .000   &   .000     &   .005    &     .000     & .001     \\
\cline{1-6}
              &  .110    &  .110    &  .110   &  .110                       \\
Fr\'ejus      &  .064    &  .007    &  .082   &  .104                       \\
              &  .000    &  .000    &  .000   &  .000                       \\
\cline{1-5}
              &    .104        &   .104    &  .100                       \\
Kamioka       &    .064        &   .069    &  .078                    \\
              &    .000        &   .000    &  .003                    \\
\cline{1-4}
              &     .082       &  .082                             \\
Hawaii        &     .064       &  .069                            \\
              &     .000       &  .000                          \\
\cline{1-3}
              &   .062                           \\
Sydney        &   .057                           \\
              &   .006                          \\
\cline{1-2}
\end{tabular}
\end{center}
\end{table}

For the special case where one of the sites is the South Pole, it is
intuitively obvious that Pyh{\"a}salmi is the best location among
the proposed ones, being farthest at north. For the most
conservative scenario, in which no new large detectors will be
built, the combination of IceCube at the South Pole plus
Super-Kamiokande detector already existing in Japan offers a 73\%
probability that a comparison between a shadowed and an un-shadowed
neutrino signal could be observed, and 17\% probability for this
comparison in the case of core shadowing. Notice that the case of a
detector in Pyh\"asalmi and another one in Hawaii, which is of
importance for geo-neutrino research, offers also a nice opportunity
for Earth effect detection in SN neutrinos. The probability that
exactly one of them is in a shadowed position exceeds 50\% (20\% for
core shadowing), whereas the probability that one or both are
shadowed is about 80\% (20\% for core shadowing).

\section{Conclusions}                          \label{sec:conclusions}

The possibility to detect Earth matter effects in the neutrino signal
from the next Galactic SN is a powerful tool to probe the neutrino
mass hierarchy.  Next-generation large volume detectors with excellent
energy resolution would offer the opportunity to detect directly the
specific signature of an energy-dependent modulation of the measured
neutrino flux. Even if current detectors could not reconstruct
directly these modulations, the comparison of the neutrino signal in a
shadowed with an unshadowed detector could allow one to detect Earth
effects. Of course, it is assumed that the location of the SN in the
sky can be determined by observations in the electromagnetic
spectrum. In the unlikely case that this is not possible, neutrinos
alone can be enough to determine the SN location with sufficient
precision~\cite{triang, Tomas:2003xn}.

Motivated by these opportunities, we have provided the first
detailed study of the probability that a detector at a given
geographic latitude will observe a Galactic SN in an Earth-shadowed
or core-shadowed position. We have shown that this probability is
rather insensitive to detailed assumptions about the uncertain
distribution of core-collapse events in the Galaxy. The main effect
is the simple geometric constraint that SNe occur within the
Galactic disk.

We find that a far-northern location such as the Pyh\"asalmi mine in
Finland, the preferred site for the LENA scintillator detector, is
almost optimal for observing a SN signal shadowed by the Earth. The
shadowing probability  is close to 60\%, against an average of 50\%
for a random location on the Earth. The shadowing probability
$P_0(\lambda)$ depends only mildly on the latitude $\lambda$ for
$\lambda>40^\circ$, a condition fulfilled by most of the northern
locations proposed for next-generation experiments. The behavior for
core shadowing is similar, although the advantage of a northern site
is more pronounced. The core-crossing probability varies by almost a
factor of two between a far southern and a far northern site, where
it reaches almost~12\%.

We have also studied the case of two detectors. It is of interest
because one may be able to compare a shadowed with an unshadowed SN
neutrino signal and thus diagnose Earth effects even if both
detectors lack the energy resolution that is necessary to observe a
modulated signal. On the other hand, if both detectors see a
shadowed signal one may perform Earth tomography because the
observed neutrinos would cross different geophysical
layers~\cite{Lindner:2002wm}. For the South Pole, where IceCube will
be completed in a few years, combined with Kamioka the probability
that at least one of them is shadowed is almost 75\%, for the South
Pole combined with Pyh\"asalmi it is almost 90\%.

One particular scenario consists of a large-volume scintillator
detector located at Pyh\"asalmi to measure the geo-neutrino flux in
a continental location and another such detector in Hawaii to
measure it in an oceanic location.  The probability that exactly one
of them is shadowed exceeds 50\% whereas the probability that one or
both are shadowed is about~80\%. Therefore, Pyh\"asalmi and Hawaii
are not only complementary for the purpose of geo-neutrino
observations, but also for observing Earth matter effects in SN
neutrinos.

The various probabilities relevant for two detectors at arbitrary
locations are difficult to represent in a useful figure or table.
Therefore, based on our benchmark Galactic SN distribution we
provide an online tool that allows one to calculate the various
Earth and core-shadowing probabilities for one or two detectors at
arbitrary geographic locations~\cite{webshadowing}.

\section*{Acknowledgments} 

We thank Lothar Oberauer for interesting discussions in the initial
stage of this work.
We thank John Beacom, Eligio Lisi, Cecilia Lunardini and Daniele
Montanino for careful reading the manuscript and for valuable
suggestions and comments. Moreover, we thank Roberto Mirizzi for
help with designing the online tool. This work was partially
supported by the European Union under the ILIAS project, contract
No.~RII3-CT-2004-506222 and by the Deutsche Forschungsgemeinschaft
under Grant No.~SFB-375. The work of A.M.\ is supported in part by
the Italian ``Istituto Nazionale di Fisica Nucleare'' (INFN) and by
the ``Ministero dell'Istruzione, Universit\`a e Ricerca'' (MIUR)
through the ``Astroparticle Physics'' research project.

\section*{References} 

\end{document}